\documentclass{article}
\newcommand{\be}{\begin{equation}}
\newcommand{\ee}{\end{equation}}
\newcommand{\bt}{\beta}
\begin{document}
\baselineskip=22pt
\hoffset=-3cm
\voffset=-3cm
{\Large FRW-TYPE UNIVERSE WITH VACUUM ENERGY DENSITY}\\
\vspace{0.6in}

{\large\bf Arbab I. Arbab\footnote{arbab64@hotmail.com}}\\
\vspace{0.1in}

{\large\it Department of Physics, Faculty of Applied Sciences,
Omdurman Ahlia University,
P.O. Box 786, Omdurman, SUDAN}

\vspace{0.6in}

We have considered a cosmological model with a cosmological
constant of the form $\Lambda=3\alpha\frac{\dot
R^2}{R^2}+\bt\frac{\ddot R}{R}\ , \ \alpha, \ \bt=\rm const.$ The
cosmological constant is found to decrease as $t^{-2}$ and the
rate of particle creation is smaller than the Steady State value.
We have found that this behavior gives
$\frac{\Lambda_{Pl}}{\Lambda_p}=10^{120}$ where $\Lambda_{Pl}$ is
the value of $\Lambda$ at Planck time. Solutions with
$\bt=3\alpha$ in the radiation dominated era and $\bt=6\alpha$ in
the matter dominated era are equivalent to the FRW results. We
have found an inflationary solution of the de-Sitter type with
$\bt=3-3\alpha$. Some problems of the Standard Model may be
resolved with the presence of the above cosmological constant in
the Einstein's equation. Since observations suggest a contribution
of the vacuum energy density in the range
$0.40<\Omega^\Lambda<0.76$, one gets $4<\beta<12$. If $\alpha=0$
the minimum age of the universe is found to be $H_p^{-1}$ ($H_p$
is the present Hubble constant) with $\beta=\infty$.
\\
\vspace{0.3cm}
\large
\\
\\
{\bf 1. Introduction}
\\
\\
Very recent results [1] suggest that the age of the universe is about
$11-14$ billion years. However results obtained from Standard Model
of cosmology are not consistent with these findings.
Numerous models have been presented to reconcile these contradictions
with
observation. Only few of these models have given a satisfactory
solution.
The presence of a cosmological constant ($\Lambda$) in a given
cosmology prolongs
the age of the universe. The inflationary paradigm requires the
Universe to
have a critical density but observations do not support this.
One of the motivations for introducing $\Lambda$ term is to reconcile
the
age parameter and the density parameter of the universe with current
observational data [15, and references therein].
In an attempt to resolve these problems with current observations,
we suggest a variation law for the cosmological constant
of the form $ \Lambda=3\alpha\frac{\dot R^2}{R^2}+\beta\frac{\ddot
R}{R}\ ,
\ \alpha, \  \bt=\rm const.$
Observational data indicate that $\Lambda\sim 10^{-55}\rm cm^{-2}$
while
particle physics prediction for $\Lambda$ is greater than this value
by
a factor of order  $10^{120}$. This discrepancy is known as the
{\it cosmological constant} problem. It is interesting that this
decay law
helps to resolve this {\it problem}.
The entropy problem that exists in the Standard Model can be solved
by
the proposed decay law. The success of the inflationary model in
extending
the Big Bang models strengthen the theoretical prejudice toward a
flat universe.
When we allow $G$ to vary with time we retain the conventional energy
conservation law, i.e., the variation of $\Lambda$ is cancelled by
the variation
of $G$. Though the proper way is to look for a field theoretic
model for the variation of $G$ and $\Lambda$ nevertheless our present
approach
could be considered as a limiting case of some  covariant theory yet
to be discovered. Brans-Dicke theory [14] is one type of these
theories but this
theory allows only a decreasing $G$ at the present time. The
variation of $G$
would have a significant effect on the evolution of stellar objects.
Thus
one way to obtain a restriction on the way in which $G$ vary is via
its impact on the evolution of astrophysical objects.

A very recent discovery suggests that the universe is flat [12]. Some
workers have

shown that the universe is accelerating. In section 6. we show that
the universe must
be accelerating if $\Lambda > 0$. This may be due to the fact that if
gravity
is increasing then the universe has to increase its expansion rate to
escape
the future collapse. Or alternatively, the decayed vacuum energy is
given
as a kinetic energy to accelerate the expansion of the universe.
In this paper we will consider a flat universe which preserves
the spirit of the inflationary scenario.
The present model could resolve many of Standard Model problems with
observation
and thus could become a viable candidate as an alternative model.
\\
\\
{\bf 2. The model}\\
\\
In a Robertson Walker metric, the Einstein's field equations
with  variable cosmological and gravitational `constants' and  a
perfect
fluid yield [2]
\be
 3\frac{\dot R^2}{R^2}+\frac{3k}{R^2}=8\pi G\rho+\Lambda\ ,
\ee
\be
 2\frac{\ddot R}{R}+\frac{\dot R^2}{R^2}+\frac{k}{R^2}=-8\pi
Gp+\Lambda\ ,
\ee
where $\rho$ is the fluid energy density and $p$ its pressure.
The equation of the state is usually given by
\be
 p=(\gamma-1)\rho\ ,
\ee
where $\gamma$ is a constant. Elimination of $\ddot{R}$  gives
\be
3(p+\rho)\dot{R}=-(\frac{\dot{G}}{G}\rho+\dot{\rho}+\frac{\dot\Lambda}{8\pi
G})R.
\ee
\\
{\bf  3.  Particle creation}\\
\\
{\it 3.1 Matter-dominated universe}(MDU)
\\
\\
For a pressure-less MDU ($ p=0$) and for a constant gravitational
constant ($G)$,
 eq.(4) reads
\be
\frac{d(\rho R^3)}{dt}=-\frac{R^3}{8\pi G}\frac{d\Lambda}{dt}\ .
\ee
In this paper we will consider a decay law of the form [3,4]
\be
\Lambda=3\alpha\frac{\dot R^2}{R^2}+\bt\frac{\ddot R}{R}\ ,
\ee
where $\alpha$ and $\bt$ are dimension-less constants.
We suggest from the linear relationship between $\Lambda$ and the
Ricci scalar
in the Einstein field equation that $\Lambda$
has a general variation which resembles this scalar. For flat space
one gets
the above variation. For a flat universe
$(k=0)$, eqs.(2) and (6) yield
\be
 (2-\bt)\ddot RR=(3\alpha-1)\dot R^2  \ ,
\ee
which can be solved to give
\be
 R(t)=[\frac{A(3-3\alpha-\bt)}{(2-\bt)}t]^{(2-\bt)/(3-3\alpha-\bt)} \
,
\ee
where $A=\rm constant$. Using eq.(8), eq.(6) becomes
\be
\Lambda(t)=\frac{(2-\bt)(6\alpha-\beta)}{(3-3\alpha-\bt)^2}\frac{1}{t^2}\
.
\ee
From eqs.(1), (6) and (8) the energy density takes the form
\be
\rho(t)=\frac{(2-\bt)}{4\pi G(3-3\alpha-\bt)}\frac{1}{t^2} \ .
\ee
The vacuum energy density ($\rho_v$) is given by
\be
\rho_v(t)=\frac{\Lambda}{8\pi G}=\frac{(2-\bt)(6\alpha-\bt)}{8\pi
G(3-3\alpha-\bt)^2}\frac{1}{t^2}\ .
\ee
The deceleration parameter ($ q$) is defined as
\be
 q=-\frac{\ddot RR}{\dot R^2}=\frac{1-3\alpha}{2-\bt}\ , \ \bt\ne2\ .
\ee
The density parameter of the universe ($\Omega^m$) is given by
\be
\Omega^m=\frac{\rho}{\rho_c}=\frac{2}{3}\frac{(3-3\alpha-\bt)}{(2-\bt)}\
\ ,\ \bt\ne2\ ,
\ee
 where $\rho_c=\frac{3H^2}{8\pi G}$ is the critical energy density of
the
 universe and $ H=\frac{\dot R}{R}$ is the Hubble constant.
We notice that the Standard Model formula $\Omega^m=2q$ is now
replaced by
$\Omega^m=\frac{2}{3}q+\frac{2}{3}$. However, both models give $
q=\frac{1}{2}$ for
a critical density.
The density parameter due to vacuum contribution is defined as
$\Omega^\Lambda=\frac{\Lambda}{3H^2}$.
Using eqs.(8) and (9) this yields
\be
\Omega^\Lambda=\frac{(6\alpha-\bt)}{3(2-\bt)}\ , \ \ \bt\ne 2\ .
\ee
We shall define $\Omega_{\rm total}$ as
\be
\Omega_{\rm total}=\Omega^m+\Omega^\Lambda\ ,\ \ and \
\rho_{total}=\rho+\rho_v\ .
\ee
Hence eqs.(13), (14) and (15) give $\Omega_{\rm total}=1$. This
situation is
favored by the inflationary scenario.
\be
 t_p=\frac{(2-\bt)}{(3-3\alpha-\bt)}H_p^{-1} ,\ \
\Omega^m_p=\frac{2}{3}\frac{(3-3\alpha-\bt)}{(2-\bt)} \ , \ \
\Lambda_p=\frac{(6\alpha-\bt)}{(2-\bt)}H_p^2\ \ , \ \ \bt\ne 2,\ \
\ee
(hereafter the subscript `p' denotes the present value of the
quantity).
For ages larger than the Standard Model one requires
$\bt<6\alpha$ and for $ t_p>0$, $\bt<2$ and $\alpha>1/3$.
This constraint indicates that $\Lambda$ is positive.
The precise value of $\alpha$ and $\bt$ has to be determined from
observational
data.\\
We now turn to calculate the rate of particle creation (annihilation)
$ n$, which is defined as [5]
\be
 n=\frac{1}{R_p^3}\frac{d(\rho R^3)}{dt}|_p\ .
\ee
Using eqs.(5), (8), (9) and (16) one obtains
\be
 n_p=\frac{(6\alpha-\bt)}{(2-\bt)}\rho_p H_p \ ,\ \ \bt\ne 2\ .
\ee
We remark that this rate is less than that of the Steady State
model ($=3\rho_0 H_p$). If $\bt=6\alpha$ then $\Lambda=0\ , \ n_p=0\
,
\ t_p=\frac{2}{3} H_p^{-1}$ and $\Omega^m_p=1$. This case is
equivalent to
the Standard Model result. We observe that when $\alpha=\frac{1}{3}$,
$q=0$
and $\beta$ decouples from all cosmological parameters.
\\
 \\
{3.2 \it Radiation-dominated universe} (RDU)
\\
\\
This is characterized by the equation of the state $
p=\frac{1}{3}\rho\ (\gamma=4/3)$.
In this case, for a flat universe $(k=0)$, eqs.(1), (2) and (6) yield
\be
 3(1-2\alpha)\frac{\dot R^2}{R^2}+(3-2\bt)\frac{\ddot R}{R}=0\ .
\ee
This can be solved to give
\be
 R=[\frac{(3-2\bt)D}{(3-3\alpha-\bt)}t]^{(3-2\bt)/2(3-3\alpha-\bt)}\
,\ D= \rm const.
\ee
For an expanding universe $\bt<3/2$ and $\alpha>1/2$.
Eqs.(1), (4), (6) and (20) give
\be
\Lambda=\frac{3(3-2\bt)(3\alpha-\bt)}{4(3-3\alpha-\bt)^2}\frac{1}{t^2}=\frac{3(3\alpha-\bt)}{(3-2\bt)}H^2\
\ ,
\ee
\be
\rho=\frac{3}{32\pi G}\frac{(3-2\bt)}{(3-3\alpha-\bt)}\frac{1}{t^2}\
,
\ee
Thus an expanding universe ($\bt<3/2, \alpha>1/2$) demands that the
cosmological constant to be positive.
It is evident that if $\bt=3\alpha$ the FRW results will be recovered
viz. $ R\propto t^{1/2},\ \rho\propto t^{-2},\ \Lambda=0\ .$
\\
\\
{\bf 4.  No particle creation}\\
\\
We now consider a model in which both $G$ and $\Lambda$ vary with
time
in such a way  the usual energy conservation law holds.
\\
{\it 4.1 Matter-dominated universe}\\
\\
Equation (4) can be split to give [2,6]
\be
\dot\rho+3 H\rho=0\ ,
\ee
and
\be
\dot\Lambda+8\pi\dot G\rho=0\ .
\ee
Using eqs.(8) and (9), eqs.(23) and (24) yield
\be
 R(t)=[\frac{A(3-3\alpha-\bt)}{(2-\bt)}t]^{(2-\bt)/(3-3\alpha-\bt)} \
,
\ee
where $A=\rm constant$.
\be
\Lambda(t)=\frac{(2-\bt)(6\alpha-\beta)}{(3-3\alpha-\bt)^2}\frac{1}{t^2}\
.
\ee
\be
\rho(t)=Ft^{-3(2-\bt)/(3-3\alpha-\bt)}\ , \ \ F=\rm const.
\ee
and
\be
 G(t)=[\frac{(2-\bt)}{4\pi
F(3-3\alpha-\bt)}]t^{(6\alpha-\bt)/(3-3\alpha-\bt)}\ \, .
\ee
Equation (24) represents a coupling between vacuum and gravity and
that the vacuum
decays to strengthen the gravitation interaction that will induce an
acceleration
of the expansion of the universe. Hence as long as gravity is
increasing the
expansion of the universe will continue. The variation of $G$ could
have been
overwhelming in the early universe. This big gravitational force
might have
been the cause for stopping the rapid expansion during inflationary
period
and later assist in making the universe matter dominated. This
because
the increasing gravity forces smaller particles to form bigger
ones.\\
For $\bt=0, \alpha=0$,  $G$=const. and $\rho=Dt^{-2}$ and $\rm
R=[\frac{3}{2}At]^{2/3}$,
which is the familiar FRW result.
Moreover, the case $\bt=6\alpha$ is equivalent to the Standard Model
result.
Clearly for $\bt<2\ , \ \alpha>1/3$ the gravitational constant
increases
with time. In an earlier work [7] we have considered the effect of
bulk
viscosity in variable $\rm G$ and $\Lambda$ models. We have shown
that many
of non-viscous models are equivalent to viscous models. The present
model
is equivalent to a viscous model with bulk viscosity $(\eta$) varying
as
$\eta\propto t^{-n/(1-n)}$ where $\eta\propto \rho^n$ with
$ n=\frac{(\frac{2}{3}\bt-\alpha-1)}{(\bt-2)}$.
It has been shown that the development of the large-scale anisotropy
is given by the ratio of the shear $\sigma$ to the volume expansion
($\theta=3\frac{\dot R}{R}$) which evolves as [8]
\be
\frac{\sigma}{\theta}\propto t^{-(3+3\alpha-2\bt)/(3-3\alpha-\bt)}\ .
\ee
The present observed isotropy of the Universe requires this
anisotropy
to be decreasing as the universe expands. Thus an increasing $G$
guarantees
an isotropic universe.
For example, if $2\bt=3+3\alpha$ then $ G\propto t^{-1}, R\propto
t^{1/3},
\rho\propto t^{-1}$. This variation of $G$ was considered by Dirac in
his Large Number Hypothesis (LNH) model [9]. According to Dirac model
one has
a constant anisotropy and this may be a problem if the universe was
not born
isotropic because the anisotropy would not decay with time.
\\
 \\
{\it 4.2 Radiation-dominated universe}
\\
\\
Equation (4) now reads
\be
\dot\rho+4H\rho=0\ ,
\ee
and
\be
\dot\Lambda+8\pi \dot G\rho=0\ .
\ee
Employing eqs.(20) and (21), eqs.(30) and (31) yield
\be
\rho=Ct^{-2(3-2\bt)/(3-3\alpha-\bt)}\ \ , \ \
G=[\frac{3(3-2\bt)}{32\pi
C(3-3\alpha-\bt)}]t^{2(3\alpha-\bt)/(3-3\alpha-\bt)}\ \ ,\ \ C=\rm
const.
\ee
We observe that when $\bt=3\alpha$ the familiar FRW model is
recovered.
Abdel Rahman [2] has recently considered a closed universe model with
a critical energy density
where both $G$ and $\Lambda$ are variable. He found that $ R\propto
t\ , G\propto t^{2}\ ,
\rho\propto t^{-4}$ in the radiation era.
His solution corresponds to $\alpha=1/2$ and a free $\bt$.
Thus both model, albeit different, evolve in the a similar way in the
early universe.
\\
The large-scale anisotropy (see eq.(29)) becomes
\be
\frac{\sigma}{\theta}\propto t^{-3(3-2\bt)/(3-3\alpha-\bt)}\propto
\rho^{3/2} \ .
\ee
Once again this anisotropy decreases with time as long as $\rho$
decreases
with expansion.
\\
 \\
{\bf 4.3 Static solutions}\\
\\
A static solution can be obtained for both matter and radiation
dominated universes with  $\bt=2$ and $\bt=3/2$, respectively.
Thus
\be
\rm R=const. , \Lambda=0\ , \ \rho_{total}=0\ , \ n=0\ .
\ee
It has been claimed by Kalligas {\it et al.} [10] that they have
obtained a
static universe with variable $\rm G$ and $\Lambda$.
In fact, their solution is nothing but the above solution, since with
$\rm R=const.$ eqs.(1) and (2) give $\Lambda=0$ so that $\rm
G$=const.
Thus their claim of a static solution with variable $\rm G$ and
$\Lambda$
can not be true with $ p\ne-\rho$.
\\
 \\
{\bf 5. An inflationary solution}  \\
  \\
This solution is obtained if we set $\rm H= const$. Thus eqs.(1) and
(2)
give $\bt=3-3\alpha$ so that $\Lambda=3H^2$. This can be integrated
to give
\be
 R=\rm const\exp(\sqrt{\Lambda/3}t)\ , \ \rho_{total}=\rho_v \ \ , \
\ G=\rm const.\ .
\ee
This is the familiar de-Sitter inflationary solution.
\\
\\
{\bf 6. An accelerating universe}\\
\\
Now consider the case $\alpha=0$, i.e., $\Lambda=\beta\frac{\ddot
R}{R}$.
Moreover, from eqs.(1)-(3) one finds
\be
\ddot R=\frac{8\pi G}{3}(1-\frac{3}{2}\gamma)\rho
R+\frac{\Lambda}{3}R.
\ee
Thus writing $\Lambda$ in the above form is equivalent to considering
the universe
to be filled with a fluid whose equation of state is given by
$p=-\frac{1}{3}\rho.$
Moreover, one may attribute the cosmic acceleration of the universe
to the decay of the
vacuum energy density (or $\Lambda$).
It follows from eq.(8)-(14) that in the matter dominated Universe
with $G$  constant
\begin{equation}
R=[\frac{A(\beta-3)}{(\beta-2)}t]^{(\beta-2)/(\beta-3)}, \ \
\Lambda=\frac{\beta(\beta-2)}{(\beta-3)^2}\frac{1}{t^2},  \ \
\rho=\frac{1}{4\pi G}\frac{(\beta-2)}{(\beta-3)}\frac{1}{t^2},\ \
\beta\ne 3,\ \beta\ne2.
\end{equation}
\begin{equation}
\rho_v=\frac{\Lambda}{8\pi G}=\frac{1}{8\pi
G}\frac{\beta(\beta-2)}{(\beta-3)^2}\frac{1}{t^2},\ \ \beta\ne 3.
\end{equation}
\begin{equation}
q_p=-\frac{\ddot RR}{\dot R^2}|_p=\frac{1}{2-\beta}, \ \
t_p=\frac{(\beta-2)}{(\beta-3)}H_p^{-1}, \ \
\Omega^m_{p}=\frac{2}{3}\frac{(\beta-3)}{(\beta-2)}, \ \ \beta\ne3, \
\beta\ne2
\end{equation}
\begin{equation}
n_p=\frac{\beta}{(\beta-2)}\rho_pH_p, \ \
\Omega_p^\Lambda=\frac{\beta}{3(\beta-2)},
\ \ \ \Lambda_p=\frac{\beta}{(\beta-2)}H_p^2,\ \ \beta\ne 2\ \
\end{equation}
If $G$ is changing with time one obtains from eqs.(28) and (29)
\be
G(t)=[\frac{(\beta-2)}{4\pi A(\beta-3)}]t^{\beta/(\beta-3)}, \ \
\frac{\sigma}{\theta}\propto t^{(3-2\beta)/(\beta-3)},\ \ \beta\ne3,
\ \beta\ne2
\ee
We observe that for $\Omega_p^\Lambda>0$, $\beta>2$ and for
$\Omega_p^m>0$, $ \beta>3$.
This implies that $q_p<0$ and thus the universe must be accelerating
at the present era.
We also notice that $G$ is an increasing function of time. Thus one
may attribute
the accelerating universe to the increasing gravity, i.e., the
universe has
to accelerate in order to overcome the ever increasing gravity.
Others  ascribed
this to a scalar field that is still rolling down its potential
``quintessence'' to justify the acceleration of the universe [13].
We see that all the cosmological parameters shown above are functions
of $\beta$.
The minimum age of the universe is $H_p^{-1}$ and corresponds
to $\beta=\infty$. We restore the Einstein-de Sitter model when
$\beta=0$.
We also note that
$\frac{\Lambda_{Pl}}{\Lambda_p}=(\frac{10^{17}}{10^{-43}})^2=10^{120}$,
where $\Lambda_{Pl}$ is the value of $\Lambda$ at Planck time. This
is evident
from eqs.(9) and (21).
This may help in solving the cosmological constant problem. Thus the
cosmological
constant is very small today because the universe is very old.
We then have a model with one parameter that can be obtained from
observation.
Recent constrains on cosmological parameters from MAXIMA-1 [11]
suggest that
$0.40<\Omega^\Lambda<0.76$, this implies that $4<\beta<12$. These
results are found to
be consistent with a flat universe. This constraint on $\beta$
imposes a
stringent constraint on  all cosmological parameters so far known.
The case $\beta=3$  corresponds to an inflationary solution in the
matter dominated universe.
If $G$ is increasing with time the presently observed isotropy of the
universe
could be one of its consequences, as is evident from eq.(41). The
case $\beta=2$
represents a static universe, which is nonsensical for the present
universe.
\\
\\
{\bf 7. Conclusion}
\\
\\
We have considered the decay law for $\Lambda$ of the form
$\Lambda=3\alpha\frac{\dot R^2}{R^2}+\bt\frac{\ddot R}{R}\ , \
\alpha,
\ \bt=\rm const.$ The cosmological consequences of the model are
shown to be very interesting. Many models in the literature can be
retrieved from this model with particular choice of $\alpha$ and
$\bt$. The exact values of these constants still require more
observational data to be determined. We note that the cosmological
problems of the standard model, namely, the age problem,
cosmological constant problem and the isotropy and homogeneity of
universe, which are still unresolved could be tackled with our
present approach. We have found that when $\alpha=0$ the minimum
age of the universe is $H_p^{-1}$(corresponds to $\beta=\infty$).
As present estimates give $\Omega_p^\Lambda=2/3$ and
$\Omega_p^m=1/3$, we suggest a model of $\alpha=0$ and $\beta=4$,
which  gives an age of $2H^{-1}_p$ and $q_p=-0.5$, as describing
our present universe. The inflationary solution, which solves the
Standard Model problems, is shown to be built in the model. A
model with $\alpha=0$ gives rise to an accelerated expansion of
the universe. This model is free from a lot of  cosmological
problems and could fit well with the present observational data.
This model contains one free parameters viz., $\beta$ that can be
obtained from the present observational data.
\\
 \\
{\bf ACKNOWLEDGMENT}\\
\\
I would like to thank the referees for  their critical comments on
the manuscript.
\\
\\
{\bf REFERENCES}
\\
\\
1- W.L. Freedman, {\it Los Alamos Preprint, astro-ph/9905222}\\
2- A.-M.M. Abdel-Rahman, {\it Gen. Rel. Gravit.}{\bf 22}(1990), 655\\
3- J.M. Overdin and F.I.Cooperstock, {\it Phys. Rev.}{\bf D58}(1998),
043506\\
4- A.S. Al-Rawaf, {\it Mod. Phys. Lett. }{\bf A 13}(1998), 429\\
5- J. Matyjasek, {\it Phys. Rev.}{\bf D51}(1995), 4154\\
6- A. Beesham, {\it Nouvo Ciment.}{\bf B96}(1986), 17,
{\it Phys. Rev.}{\bf D48}(1993), 3539\\
7- A.I. Arbab, {\it Gen. Rel. Gravit.}{\bf29}(1997), 61\\
8- J.D. Barrow, {\it Mon. Not. astr. Soc.}{\bf 184}(1978), 677\\
9- P.A.M. Dirac, {\it Nature}{\bf 139}(1937), 323\\
10- D. Kalligas, P.Wesson and C.W.F.Everitt, {\it Gen. Rel.
Gravit.}{\bf 24}(1992),351\\
11- A. Balbi {\it et al}, {\it astro-ph/0005124}(2000)\\
12- P.de Bernardis {\it et al}, {\it astro-ph/0004404}(2000)\\
13- P.J.E. Peebles and B.Ratra, {\it Astrophys. J.}{\bf 325}(1988),
L17\\
14- C. Brans. and R.H. Dicke, {\it Phys. Rev.}{\bf 124}(1961), 925\\
15- M. \"{O}zer, {\it  Astrophys.J.}{\bf 520}(1999), 45\\
\end{document}